\documentclass[a4paper,twocolumn,english,prl,superscriptaddress,showpacs,longbibliography,fixfloat,longbibliography]{revtex4-1}
\usepackage[latin9]{inputenc}
\usepackage{amsmath}
\usepackage{amssymb}
\usepackage{graphicx}

\makeatletter

\pdfpageheight\paperheight
\pdfpagewidth\paperwidth

\usepackage{xcolor}
\definecolor{darkblue}{rgb}{0.1,0.2,0.6} 
\definecolor{lightblue}{rgb}{0.1,0.1,1.0}
\definecolor{darkred}{rgb}{0.8,0.1,0.2}
\usepackage[colorlinks,citecolor=lightblue,linkcolor=darkblue,urlcolor=lightblue] {hyperref}
\renewcommand{\BibitemShut}[1]{}

\makeatother

\usepackage{babel}
\begin{document}
\global\long\def\E{\mathrm{e}}%
\global\long\def\D{\mathrm{d}}%
\global\long\def\I{\mathrm{i}}%
\global\long\def\mat#1{\mathsf{#1}}%
\global\long\def\vec#1{\mathsf{#1}}%
\global\long\def\cf{\textit{cf.}}%
\global\long\def\ie{\textit{i.e.}}%
\global\long\def\eg{\textit{e.g.}}%
\global\long\def\vs{\textit{vs.}}%
 
\global\long\def\ket#1{\left|#1\right\rangle }%

\global\long\def\etal{\textit{et al.}}%
\global\long\def\tr{\text{Tr}\,}%
 
\global\long\def\im{\text{Im}\,}%
 
\global\long\def\re{\text{Re}\,}%
 
\global\long\def\bra#1{\left\langle #1\right|}%
 
\global\long\def\braket#1#2{\left.\left\langle #1\right|#2\right\rangle }%
 
\global\long\def\obracket#1#2#3{\left\langle #1\right|#2\left|#3\right\rangle }%
 
\global\long\def\proj#1#2{\left.\left.\left|#1\right\rangle \right\langle #2\right|}%

\title{Is there slow particle transport in the MBL phase?}
\author{David J. Luitz}
\affiliation{Max Planck Institute for the Physics of Complex Systems, Noethnitzer
Str. 38, Dresden, Germany}
\email{dluitz@pks.mpg.de}

\author{Yevgeny Bar Lev}
\affiliation{Department of Physics, Ben-Gurion University of the Negev, Beer-Sheva
84105, Israel}
\email{ybarlev@bgu.ac.il}

\date{\today}
\begin{abstract}
We analyze the saturation value of the bipartite entanglement and
number entropy starting from a random product state deep in the MBL
phase. By studying the probability distributions of these entropies
we find that the growth of the saturation value of the entanglement
entropy stems from a significant reshuffling of the weight in the
probability distributions from the bulk to the exponential tails.
In contrast, the probability distributions of the saturation value
of the number entropy are converged with system size, and exhibit
a sharp cut-off for values of the number entropy which correspond
to one particle fluctuating across the boundary between the two halves
of the system. Our results therefore rule out slow particle transport
deep in the MBL phase and confirm that the slow entanglement entropy
production stems uniquely from configurational entanglement.
\end{abstract}
\maketitle
\textbf{Introduction}.---Generic interacting quantum many-body systems
are expected to thermalize after a quench by virtue of the eigenstate
thermalization hypothesis \citep{feingold_ergodicity_1984,deutsch_quantum_1991,srednicki_chaos_1994,srednicki_approach_1999,rigol_thermalization_2008,dalessio_quantum_2016,borgonovi_quantum_2016}.
However the addition of sufficiently strong quenched disorder allows
such systems to avoid thermalization \citep{basko_metal-insulator_2006,huse_phenomenology_2014,luitz_ergodic_2017,abanin_recent_2017,alet_many-body_2018,abanin_colloquium_2019},
a phenomenon which is called many-body localization (MBL). For one
dimensional systems the existence of the MBL phase at strong disorder
is now firmly established \citep{imbrie_many-body_2016}, but the
situation for higher dimensional systems is still an open question
\citep{wahl_signatures_2019,theveniaut_many-body_2019,choi_exploring_2016,chertkov_numerical_2020}.
When all many-body eigenstates are localized, the phenomenology of
MBL is understood by the emergence of local conserved quantities called
the l-bits \citep{serbyn_local_2013,huse_phenomenology_2014,imbrie_local_2017}.
The existence of l-bits predicts the absence of particle transport
and thermalization in the MBL phase, but also an unbounded logarithmic
growth of the bipartite entanglement entropy \citep{chiara_entanglement_2006,znidaric_many-body_2008,bardarson_unbounded_2012,serbyn_local_2013,nanduri_entanglement_2014,huse_phenomenology_2014,znidaric_entanglement_2018}
up to a nonthermal, extensive, saturation value. This behavior is
in stark contrast with the thermal phase, where the entanglement entropy
grows as a power-law in time \citep{luitz_extended_2016,lezama_power-law_2019}.
Without disorder, a linear growth of the entanglement entropy is typically
observed \citep{kim_ballistic_2013}.

Within the l-bit model the production of entanglement deep in the
MBL phase does not rely on particle transport, since due to the emergent
conservation laws the particle number in any part of the system is
essentially constant for all times \citep{hauschild_domain-wall_2016}.
This view was challenged very recently in a study of the entropy of
the subsystem particle number distribution \citep{kiefer-emmanouilidis_evidence_2020},
where a long growth regime of the number entropy was observed, which
was argued to continue indefinitely in the thermodynamic limit in
the MBL phase, suggesting that there is very slow transport in the
MBL phase.

In the present work, we address this question by a detailed statistical
analysis of the behavior of the saturation values of the entanglement
and the entropy of the subsystem particle number distribution (number
entropy). We find that the fluctuations of the particle number are
strictly limited at strong disorder and preclude an indefinite growth
of the number entropy.

\textbf{Model and method}.---We consider the standard model of many-body
localization, an open spin chain with random fields: 
\begin{equation}
\hat{H}=J\sum_{i=1}^{L-1}\boldsymbol{S}_{i}\cdot\boldsymbol{S}_{i+1}+\sum_{i=1}^{L}h_{i}\hat{S}_{i}^{z},\label{eq:hamiltonian}
\end{equation}
where $J$ corresponds to the interaction between the spins and the
random fields are drawn from a box distribution with $h_{i}\in\left[-W,W\right]$.
Without loss of generality we set $J=1$, throughout this work. Using
the Jordan-Wigner transformation, this model maps exactly to a model
of interacting spinless fermions,
\begin{align}
\hat{H} & =\frac{J}{2}\left(\hat{c}_{i}^{\dagger}\hat{c}_{i+1}+\hat{c}_{i+1}^{\dagger}\hat{c}_{i}\right)+J\sum_{i=1}^{L-1}\left(\hat{n}_{i}-\frac{1}{2}\right)\left(\hat{n}_{i+1}-\frac{1}{2}\right)\nonumber \\
 & +\sum_{i=1}^{L}h_{i}\left(\hat{n}_{i}-\frac{1}{2}\right),\label{eq:spinless-fermions}
\end{align}
where $\hat{c}_{i}^{\dagger}$ creates a fermion at site $i$ and
$\hat{n}_{i}=\hat{c}_{i}^{\dagger}\hat{c}_{i}$. The model conserves
the total magnetization (respectively, the particle number), and throughout
this work we fix $\sum_{i}\hat{S}_{i}^{z}=0$, (respectively, half-filling).
While the model (\ref{eq:hamiltonian}) has been studied in great
detail, the critical disorder is only known with a large margin of
error, $W_{c}=3.8\pm1$ \citep{znidaric_many-body_2008,pal_many-body_2010,luitz_many-body_2015,devakul_early_2015,doggen_many-body_2018}.
We therefore focus on quite strong disorder strengths to be sufficiently
far from the critical regime, which is known to exhibit strong finite
size effects \citep{khemani_critical_2017,weiner_slow_2019,luitz_multifractality_2020}.

We study the behavior of the system at effectively infinite times
using full diagonalization of the Hamiltonian \eqref{eq:hamiltonian}.
We extract eigenstates and eigenvalues to evolve the initial product
states $\ket{\sigma_{1},\sigma_{2},\dots,\sigma_{L}}$ in time. Here,
$\sigma_{i}=\pm\frac{1}{2}$ are the eigenvalues of the corresponding
local $\hat{S}_{i}^{z}$ operators. The initial states have a definite
number of up spins (i.e. $\sigma_{i}=+1/2$) in any subsystem, which
corresponds to a definite number of particles in the equivalent spinless
fermion model (\ref{eq:spinless-fermions}).

\textbf{Results}\emph{.---}We consider a quench from a product state
$\ket{\psi_{0}}=\ket{\sigma_{1},\sigma_{2},\dots\sigma_{L}}$ in the
$\hat{S}_{z}$ basis, and cut the system into two subsystems of equal
size, $A$ and $B$, where spins $i=1,\dots,L/2$ are in subsystem
$A$ and spins $L/2+1,\dots,L$ are in subsystem $B$. Due to the
conservation law $\sum_{i}\hat{S}_{i}^{z}=0$, the reduced density
matrix $\hat{\rho}_{A}\left(t\right)=\tr_{B}\ket{\psi\left(t\right)}\bra{\psi\left(t\right)}$
of the subsystem A is block diagonal with blocks labeled by the number
of up spins $n_{A}$ in the subsystem. The probability $p\left(n_{A}\right)$
to find $n_{A}$ up spins (corresponding to particles in the spinless
system) in subsystem $A$ is given by the trace of the reduced density
matrix $\hat{\rho}_{A}$ in this block.

The entanglement entropy is given by,
\begin{equation}
S_{E}=-\tr\left(\hat{\rho}_{A}\ln\hat{\rho}_{A}\right)\label{eq:EE1}
\end{equation}
and the number entropy $S_{N}$ is the Shannon entropy of the number
distribution $p\left(n_{A}\right)$ \citep{lukin_probing_2019} 
\begin{equation}
S_{N}=-\sum_{n_{A}}p\left(n_{A}\right)\ln p\left(n_{A}\right).\label{eq:NE1}
\end{equation}
In the context of MBL it is useful to split the contributions to entanglement
from particle number fluctuations and the configurations of the particles,
by introducing the configurational entropy $S_{C}$, which is the
difference $S_{E}-S_{N}$ \citep{lukin_probing_2019,kiefer-emmanouilidis_evidence_2020}.

\begin{figure}
\centering \includegraphics{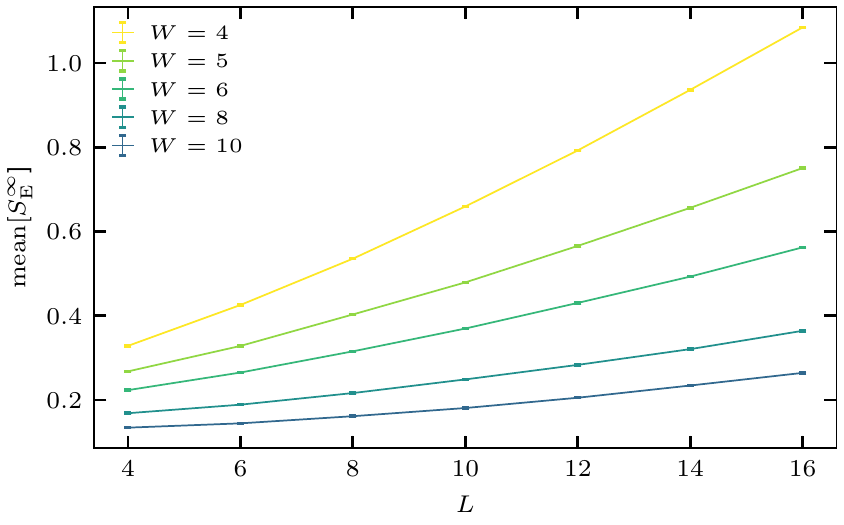} \caption{Disorder average of the long time-averaged entanglement entropy $S_{E}^{\infty}$
as a function of system size $L$ for different disorder strengths
$W$.}
\label{fig:meanEE}
\end{figure}

\begin{figure}
\centering \includegraphics{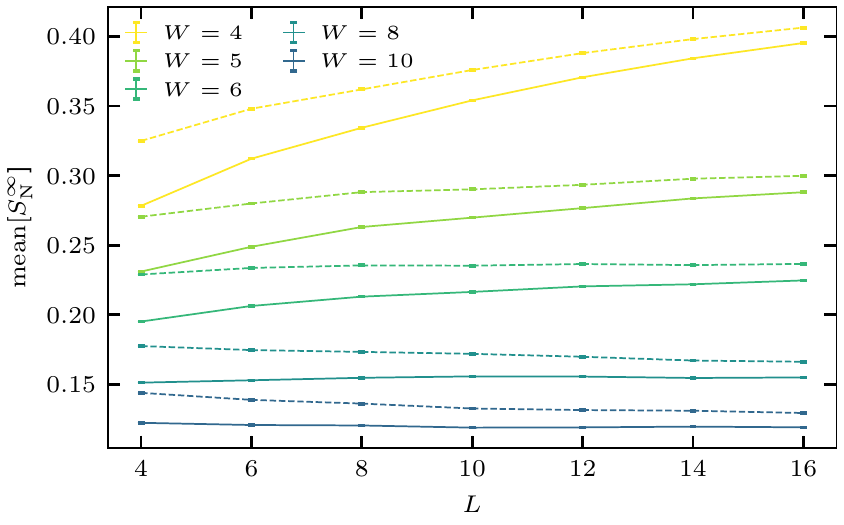} \caption{Full lines: Disorder average of the time-averaged number entropy $S_{N}^{\infty}$
as a function of system size $L$ for different disorder strengths
$W$. Dashed lines: Disorder average of the entropy of the infinite
time averaged subsystem number distribution $S\left[p^{\infty}\left(n_{A}\right)\right]$,
which is an upper bound of $S_{N}^{\infty}$.}
\label{fig:meanNE}
\end{figure}

The wavefunction $\ket{\psi(t)}$ of the system evolves according
to, 
\begin{equation}
\ket{\psi(t)}=\sum_{n}\E^{-\I E_{n}t}\braket n{\psi\left(0\right)}\ket n,\label{eq:psit}
\end{equation}
where $E_{n}$ and $\ket n$ are the eigenvalues and eigenstates of
the Hamiltonian. At any point in time we can calculate the entanglement
entropy and the number entropy $S_{N/E}\left(t\right)$, using (\ref{eq:EE1}),
(\ref{eq:NE1}) and the definition of $\hat{\rho}_{A}$. Since in
this work we are interested in the saturation values of the entropies,
we consider the infinite time-average
\begin{equation}
S_{N/E}^{\infty}=\lim_{T\to\infty}\frac{1}{T}\int\limits _{0}^{T}\mathrm{d}t\,S_{N/E}(t),\label{eq:Sinf}
\end{equation}
which is estimated numerically by averaging the entropies over $40$
time points at very late times to represent the saturation value of
the entropy for finite systems. Practically, we have checked that
the saturation values are robustly reached at times as late as $t\in\left[10^{16},10^{24}\right]$
for all sizes and disorder strengths which we consider in this work.

The entropies exhibit significant temporal fluctuations at late times
(cf. Appendix) , therefore in order to get a more robust insight of
the late time behavior, we also consider the infinite time-average
of the number distribution 
\begin{equation}
p^{\infty}\left(n_{A}\right)=\lim_{T\to\infty}\frac{1}{T}\int\limits _{0}^{T}\mathrm{d}t\,p\left(n_{A},t\right),\label{eq:pinf}
\end{equation}
which unlike $S\left[p^{\infty}\left(n_{A}\right)\right]$ is not
affected by temporal fluctuations, since it can be calculated numerically
exactly. It is easy to show that the entropy $S\left[p^{\infty}\left(n_{A}\right)\right]$
bounds from above the infinite time-averaged number entropy, $S_{N}^{\infty}$,
noting that the entropy is a concave function of $p\left(n\right)$
and using Jensen's inequality.

In this work we do not study the temporal dependence, but focus directly
on the saturation values of the entropies. If the saturation value
grows with system size, then the dynamical growth regime continues
indefinitely in the thermodynamic limit, and if is independent of
the system size, the temporal growth regime is transient.

In Fig. \ref{fig:meanEE}, we show the disorder averaged saturation
value of the entanglement entropy $S_{E}^{\infty}$, obtained by time
evolution of the wavefunction to very long times $t\geq10^{16}$ and
averaging over $40$ time points for each of the $50$ random initial
product states in addition to averaging over the disorder realizations.
We note that it is important to average over a very large number $\left(n=50\,000\right)$
of disorder realizations to obtain converged statistical averages.

It is clearly visible that the saturation value of the entanglement
entropy grows with system size. Interestingly, we observe a significant
upturn of the curves even for very strong disorder, which is only
weakly visible in previous data obtained at weak interaction strengths
\citep{bardarson_unbounded_2012,serbyn_universal_2013}.

The full lines in Fig. \ref{fig:meanNE} show the time averaged number
entropy $S_{N}^{\infty}$ as a function of system size $L$ and for
different disorder strengths $W$, spanning both the critical and
MBL regimes. While for $W\leq6$, $S_{N}^{\infty}$ still grows slightly
with system size, we observe a saturation and even a weak decrease
for strong disorders $\left(W\geq8\right)$. This becomes even more
apparent if we consider $S\left[p^{\infty}\left(n_{A}\right)\right]$,
which satisfies $S_{N}^{\infty}\leq S\left[p^{\infty}\left(n_{A}\right)\right]$
and is plotted as the dashed lines in Fig.~\ref{fig:meanNE}. We
see that $S\left[p^{\infty}\left(n_{A}\right)\right]$ is saturated
for $W=6$ and slightly decreases with system size for stronger disorders.
Since $S_{N}^{\infty}$ can not exceed $S\left[p^{\infty}\left(n_{A}\right)\right]$,
we conclude that $S_{N}^{\infty}$ is independent of the system size
for strong disorder.

\begin{figure}
\centering \includegraphics{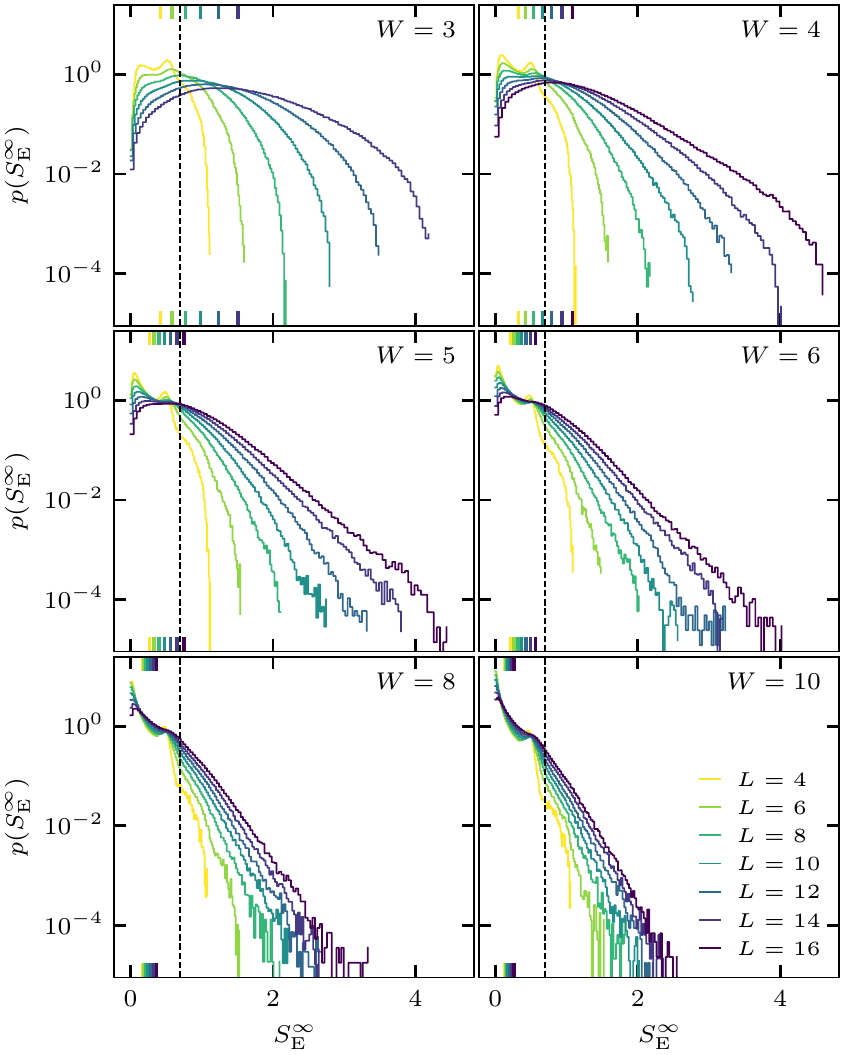} \caption{Distribution of the time averaged entanglement entropy $S_{E}^{\infty}$
for disorder strengths $W=3,4,5,6,8,10$ and different system sizes
$L$. The distributions are taken over $50\,000$ disorder realizations
and $50$ initial random product states. For each disorder realization
the time average is computed by sampling $40$ time points for $t\protect\geq10^{16}$.
The black dashed line shows the entropy $S_{E}=\ln(2)$ and the colored
ticks on the horizontal axis mark the corresponding mean of $S_{E}^{\infty}$.}
\label{fig:histEE}
\end{figure}

As the mean contains only limited information about the probability
distribution, we study the full distributions of $S_{E/N}^{\infty}$.
Fig.~\ref{fig:histEE} shows the distribution of $S_{E}^{\infty}$
over $50$ random initial product states and $50\,000$ disorder realizations.
For intermediate disorder strengths, in the critical regime $W=3,4,5$
for our system sizes, the distributions of the entanglement entropy
are very broad and the growth of the mean is clearly visible in a
reshuffling of the weight from low to high entropy as the system size
is increased. It is interesting to see that the low entropy part of
the distribution of $S_{E}^{\infty}$ is significantly different from
the distribution of the entanglement entropy of eigenstates (cf. e.g.
Fig. 10 c) and d) in Ref.~\citep{luitz_long_2016}). While the mean
entanglement of eigenstates does \emph{not} depend on the system size
\citep{yu_bimodal_2016}, $S_{E}^{\infty}$ grows with size as shown
in Fig.~\ref{fig:meanEE}, due to dephasing between the various eigenstates
which are spanning the initial state.

\begin{figure}
\centering \includegraphics{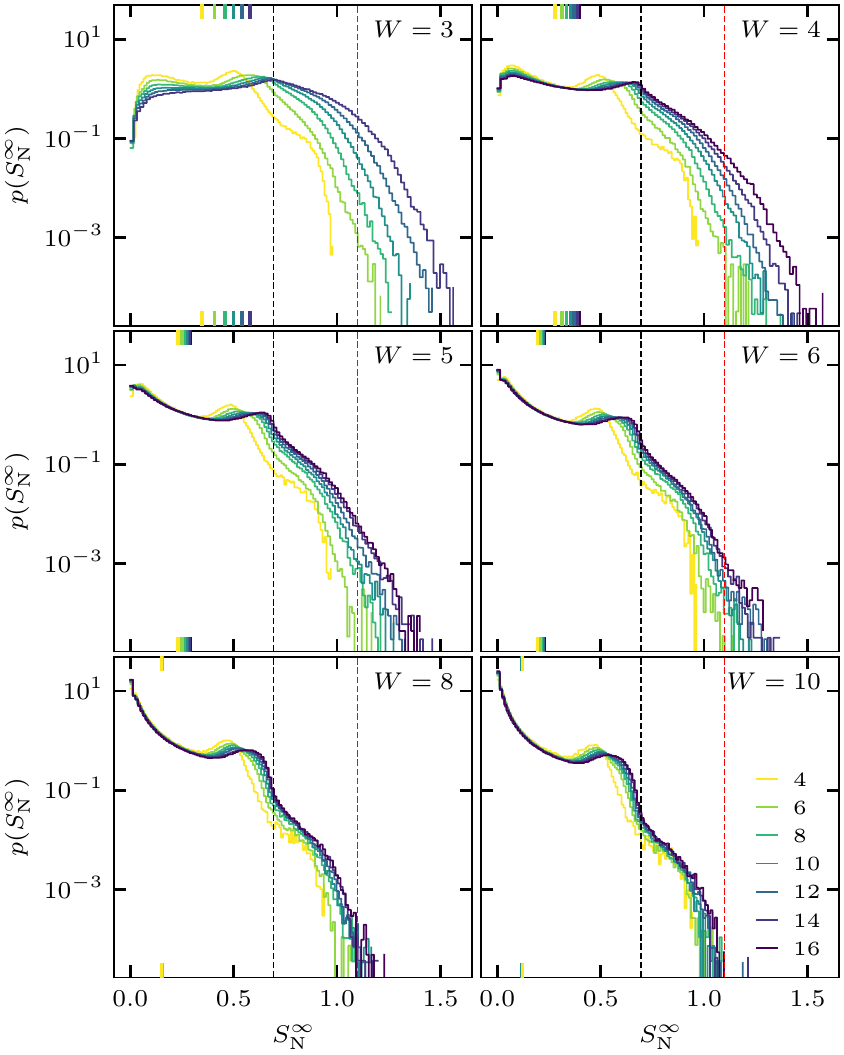} \caption{Similarly to Fig.~\ref{fig:histEE}, but for the time averaged number
entropy $S_{N}^{\infty}$. The red dashed line shows the entropy $S=\ln(3)$.}
\label{fig:histNE}
\end{figure}

At very strong disorder, the weight around $S_{E}=0$ is visibly decreasing
with increasing systems size, which leads to a corresponding increase
in the mean. For high entropies, the distribution exhibits a long,
seemingly exponential tail, with a negligible contribution to the
mean.

In Fig. \ref{fig:histNE} we consider the distribution of the time
averaged number entropy $S_{N}^{\infty}$. While in the critical regime
$\left(W=3,4,5\right)$, there is a significant reshuffling of the
weight from low to high number entropies as the system size is increased,
the entire distribution seemingly converges to a limiting distribution
at large sizes and strong disorderes $W=8,10$. This is accompanied
with an effective independence of the mean of the distribution on
the system size, as is also shown in Fig.~\ref{fig:meanNE}. For
strong disorder, the distributions exhibit a secondary peak for $S<\ln\left(2\right)$,
which is reminiscent of the $\ln(2)$ peak in the distributions of
the eigenstate entanglement entropy \citep{bauer_area_2013,luitz_long_2016,lim_many-body_2016}.
Here this peak is broadened and stays strictly below $\ln\left(2\right)$
(black dashed horizontal line) for all considered system sizes. The
observed probability distribution seems to decay significantly for
entropies larger than $\ln\left(3\right)$ (red dashed horizontal
line) as we will discuss in more detail below.

\begin{figure}
\centering \includegraphics{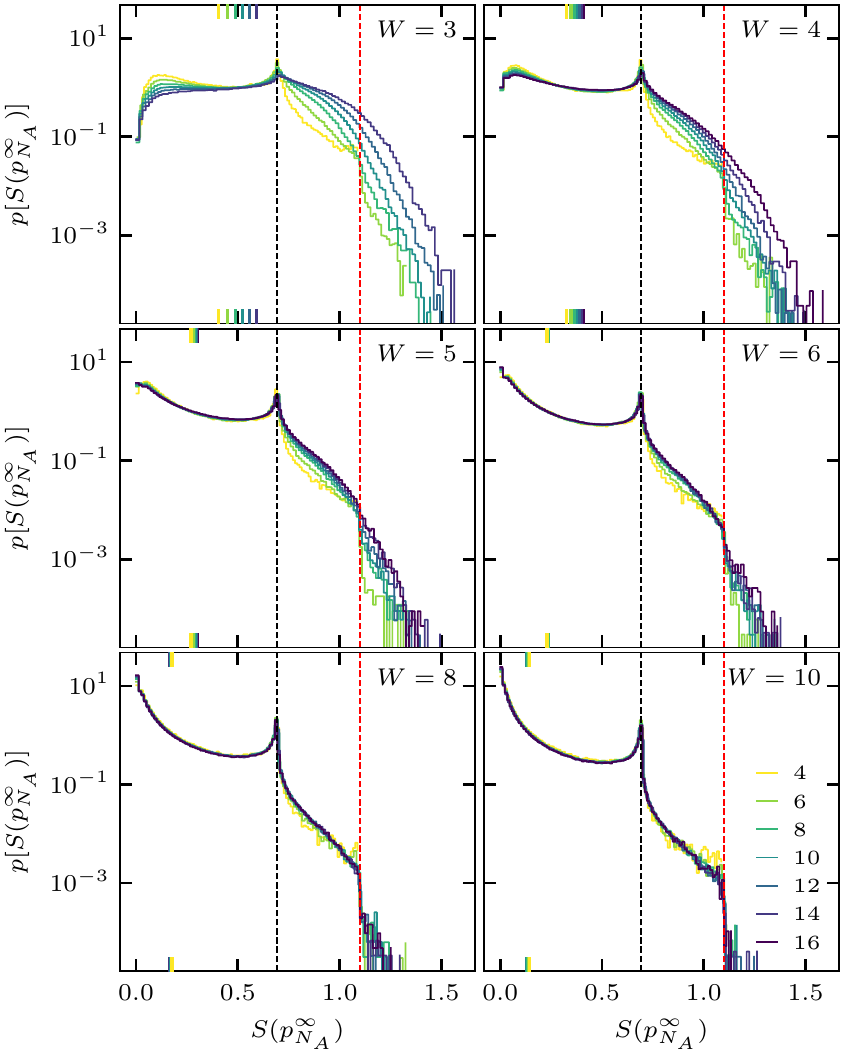} \caption{Similarly to Fig.~\ref{fig:histNE}, but for the entropy of the time
averaged subsystem particle number distribution $S\left[p^{\infty}\left(n_{A}\right)\right]$.}
\label{fig:histNElimit}
\end{figure}

For a better, quantitative understanding of the high entropy part
of the distribution of $S_{N}^{\infty}$, we consider next the distributions
of $S\left[p^{\infty}\left(n_{A}\right)\right]$, which is cleaner
due to the fact that the infinite time-average can be calculated exactly
and since it provides an upper bound to $S_{N}^{\infty}$. Fig.~\ref{fig:histNElimit}
shows the distribution of $S\left[p^{\infty}\left(n_{A}\right)\right]$,
which exhibit a sharp secondary peak at $\ln\left(2\right)$, corresponding
to an equal probability to have $n_{A},n_{A}+1$ or $n_{A,}n_{A}-1$
particles. Even more interestingly, we observe a steep decrease in
the probability density at $\ln\left(3\right)$ (red dashed line),
which correspond to an equal probability to measure $n_{A}$, $n_{A}+1$
or $n_{A}-1$ where $n_{A}$ is the particle number in the initial
state. This means that in this rare state one particle has crossed
the boundary between the two subsystems.

For strong disorder $\left(W=8,10\right)$, we see that the distribution
is completely converged with the system size, exhibiting a sharp cut-off
at $\ln\left(3\right)$. The mean of this distribution does not grow
with the system size.

\textbf{Discussion}.---In this work we presented a detailed study
of the saturation value of entanglement and number entropies including
the probability distributions over the initial product states and
disorder realizations. We have shown that at strong disorder the mean
of the saturation value of the entanglement entropy grows with system
size, which is consistent with previous literature. The distributions
of these quantities are quite broad and require a very large number
of disorder realizations to be sampled precisely. We identify that
the growth of the mean with system size results from a reshuffling
of weight from low to high entropies.

The entanglement entropy can be decomposed into a sum of the number
entropy and the configurational entropy. We show that at strong disorder
$W=8,10$ the mean of the saturation value of the number entropy does
not grow with system sizes, moreover, for large systems the entire
probability distribution converges to a limiting distribution. We
further study the entropy of the infinite-time averaged number distributions,
which allows us to show that the number fluctuations are typically
bounded by a change of one particle across the two halves of the system.
This leads to a sharp cut-off in the probability distribution of the
number entropy at $\ln(3)$. Even for the very large number of disorder
realizations used in this work, we did not observe realizations with
number entropies which exceed this limit, which leads us to conclude
that there is no particle transport for sufficiently strong disorder
and the observed growth of the number entropy in time in Ref.~\citep{kiefer-emmanouilidis_evidence_2020}
is pertinent to the critical regime and likely disappears for stronger
disorder or larger system sizes.
\begin{acknowledgments}
We are grateful to Achilleas Lazarides and Roderich Moessner for useful
comments and thank Jesko Sirker for discussions. We acknowledge financial
support from the Deutsche Forschungsgemeinschaft through SFB 1143
(project-id 247310070). YBL acknowledges support by the Israel Science
Foundation (grants No. 527/19 and 218/19).
\end{acknowledgments}

\bibliography{mbl_number_entropy}

\clearpage{}

\appendix

\section{Distribution of the configurational entropy}

\begin{figure}[h]
\centering \includegraphics{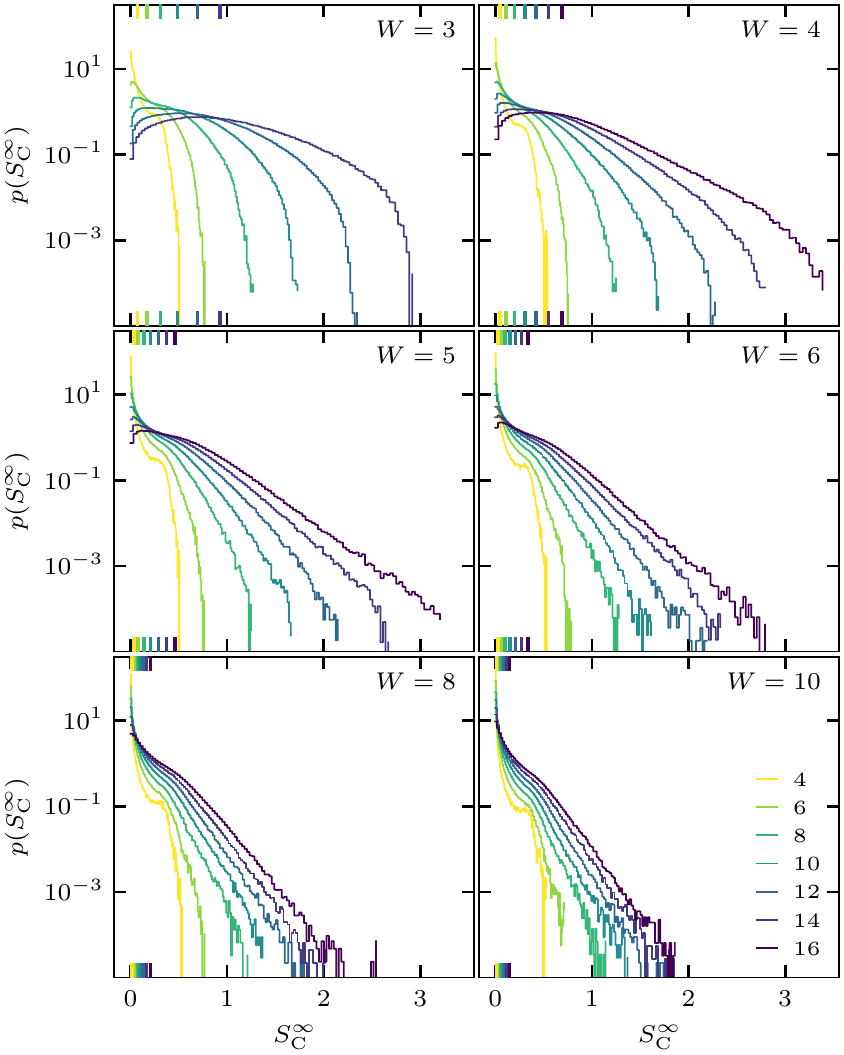} \caption{Distribution of the saturation value of the configurational entropy
for disorder strengths $W=3,4,5,6,8,10$ and different system sizes
$L$. Distributions are taken over $50\,000$ disorder realizations
and $50$ initial random product states. In each disorder realization,
the time-average is sampled over $40$ time points for $t\protect\geq10^{16}$.}
\label{fig:histCE}
\end{figure}

In this appendix, we provide additional data for the distribution
of saturation value of the configurational entropy $S_{C}^{\infty}$
, which is defined as,
\begin{equation}
S_{C}^{\infty}=S_{E}^{\infty}-S_{N}^{\infty}.\label{eq:CE}
\end{equation}

We show the full probability distribution (with its mean indicated
by colored ticks at the top and bottom of each panel) in Fig.~\ref{fig:histCE}.
The mean of the distribution grows with system size for all disorder
strengths with a slower growth at strong disorder. This growth stems
from reshuffling of the weight from low to high entropies.

The distribution of the configurational entropy is quite similar to
that of the entanglement entropy and exhibits an exponential tail
as well as a peak at zero entropy which decreases with system size.
The sharp peak located at $S_{E}^{\infty}<\ln(2)$ for small systems
in the entanglement entropy distribution is strongly suppressed in
the number entropy distribution, indicating that it stems from the
number entropy.

\section{Temporal fluctuations of the saturation value of the entanglement
and number entropies}

Besides the time average of the late time entropies, we consider also
the temporal fluctuations around the saturation value of the entropies.

\begin{figure}[h]
\centering \includegraphics{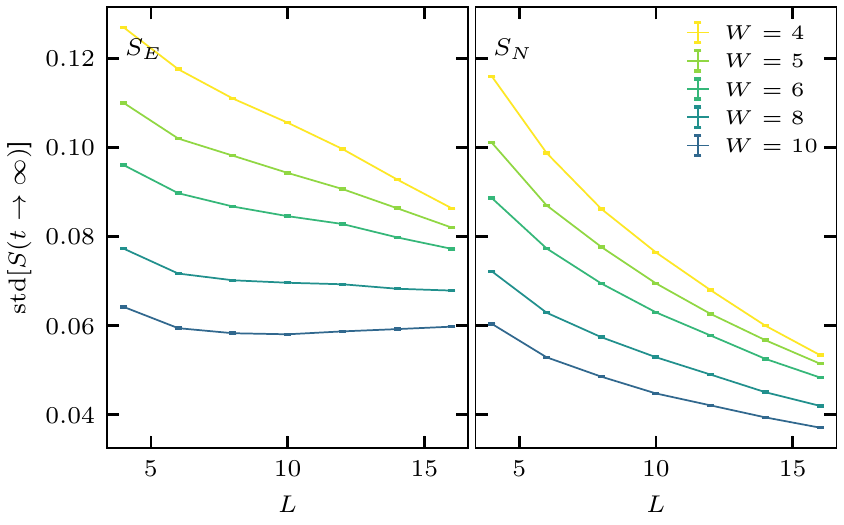} \caption{Temporal fluctuations of the entanglement entropy $S_{E}$ (left)
and the number entropy $S_{N}$ (right), as a function of the system
size and for various disorder strengths.}
\label{fig:std}
\end{figure}

While at strong disorder, the fluctuations are generally suppressed,
it is interesting to see that the fluctuations of the entanglement
entropy are larger and do not decay with system size, while the number
entropy fluctuations decrease with systems size for all disorder strengths.
\end{document}